\documentclass[prb,aps,amsmath,amssymb,prb,twocolumn,showpacs]{revtex4}
\usepackage{bm}
\usepackage{graphicx}

\begin{document}

\title{Quantum Creep and Variable Range Hopping of One-dimensional
  Interacting Electrons}

\author{Sergey V.~Malinin$^{1,2}$, Thomas Nattermann$^1$, and Bernd Rosenow$^1$}
\affiliation{$^1$Institut f\"ur Theoretische Physik, Universit\"at zu K\"oln,
50937 K\"oln, Germany\\
$^2$L.D.~Landau Institute for Theoretical Physics, Russian Academy of Sciences,
117334 Moscow, Russia }
\date{\today}


\begin{abstract}
The variable range hopping results for noninteracting electrons of
Mott and Shklovskii   are generalized to 1D disordered charge
density waves and Luttinger liquids using an instanton approach.
Following a recent paper by Nattermann, Giamarchi and Le Doussal
[Phys. Rev. Lett. {\bf 91}, 56603 (2003)] we calculate the quantum
creep of charges at zero temperature and the linear conductivity at
finite temperatures for these systems. The hopping conductivity for
the short range interacting electrons acquires the same form as for
noninteracting particles if the one-particle density of states is
replaced by the compressibility. In the present paper we extend the
calculation to dissipative systems and give a discussion of the
physics after the particles materialize behind the tunneling
barrier. It turns out that dissipation is crucial for tunneling to
happen. Contrary to pure systems the new metastable state does not
propagate through the system but is restricted to a region of the
size of the tunneling region. This corresponds to the hopping of an
integer number of charges over a finite distance. A global current
results only if tunneling events fill the whole sample. We argue
that rare events of extra low tunneling probability are not relevant
for realistic systems of finite length. Finally we show that  an
additional Coulomb interaction only leads to small logarithmic
corrections.
\end{abstract}

\pacs{
72.10.-d, 71.10.Pm, 71.45.Lr}
\maketitle


\section{Introduction}

In one dimensional systems, the interplay of interaction and
disorder gives rise to a variety of interesting physical phenomena
\cite{Giamarchi,Gruner,Giamarchi+Schulz,KaneFisher,FuNa93,YuGlMa94,ll78,Finkelstein83}.
In and below two dimensions, disorder leads to localization of all
electronic states
\cite{mt61,BeKi94,Berezinskii73,Abrahams+79,OpWe79,VoWo80}.
Interactions lead to a breakdown of the Fermi liquid concept in
one dimension and to the emergence of a Luttinger liquid instead
\cite{Haldane,KaneFisher,Giamarchi}, which is characterized by
collective excitations. In 1d electron systems, these strong
perturbations compete with each other. In the presence of a
repulsive interaction, disorder is a relevant perturbation and
drives the system into a localized phase
\cite{Giamarchi+Schulz,BeKi94}. As a result the linear
conductivity vanishes at zero temperature. Thermal fluctuations
destroy the localized phase \cite{Glatz+Natter02,Glatz+Natter04}.
At sufficiently high temperatures, i.e. if the thermal de Broglie
wavelength of the collective excitations becomes shorter than the
zero temperature localization length, the linear conductivity
exhibits a power law dependence on temperature with interaction
dependent exponents \cite{Giamarchi+Schulz}. At very low
temperatures noninteracting electrons show  hopping conductivity
\cite{Mott69,ahl71,Shklovskii_Efros_Book}. In the present paper we
study both, the nonlinear conductivity due to strong electric
fields in the zero temperature localized phase as well as the
finite temperature hopping conductivity  for interacting systems.

Interaction effects in 1d electron systems can be described by the
method of bosonization \cite{Haldane}, which directly describes the
collective degrees of freedom associated with the oscillations of a
string.  Due to the quantum nature of the system, the string can be
viewed as a two dimensional elastic manifold in a three dimensional
embedding space.  In the localized phase, this manifold is pinned by
the impurity potential.  For a classical pinned manifold, thermally
activated hopping over barriers leads to thermal creep, i.e. a
nonlinear current-voltage characteristic \cite{IoVi87,Na87}. Due to
the quantum nature of the bosonized electron system, thermal hopping
is replaced by quantum tunneling giving rise to quantum creep \cite{ngd03}.

From a quantum mechanical point of view, tunneling driven by a
potential gradient corresponds to the decay of a metastable state.
Using a field theoretical language, this decay is described by
instantons \cite{langer,Schulman,tHooft,Coleman}, i.e. solutions of
the Euclidean saddle point equations with a finite action. In the
presence of a periodic potential, instantons relate ground states
which differ by a multiple of $2\pi$. In this way, the nucleation of
a new ground state takes place. The instanton picture was used by
Maki \cite{Maki} to study the nonlinear conductivity of the quantum
sine Gordon model. Recently, this framework was used to describe
quantum creep in a sine Gordon model with random phase shifts
\cite{ngd03}, which models the physics of disordered and short range
interacting 1d electrons in the localized phase.

In the quantum sine Gordon model, instanton formation is followed by
materialization of a kink particle. The materialization is
characterized by a rapid expansion of the instanton in both space and
time. During this expansion process of the new ground state, the kink
particle moves according to its equation of motion\cite{Maki}.  Within
the Euclidean action approach, the instanton expansion is described by
an unstable Gaussian integral over the saddle point, which gives rise
to an imaginary part of the partition function describing the
instability of the metastable state \cite{langer}.  In the disordered case,
instanton formation describes tunneling from an occupied site close to
the Fermi surface to a free site close to it. However, spatial motion
of the kink particle after its materialization is impeded by new
barriers.

The central point of our paper is a discussion of the
materialization process in the disordered case: an extension of the
instanton is only possible in time direction due to the presence of
barriers in space direction. The inclusion of dissipation
\cite{feynman,CaldeiraLeggett,Hida+Eckern,Kleinert,Weiss99} is
essential for this process to occur.  A dissipative term in the
action is essential for finding a saddle point associated with an
unstable Gaussian integral on the one hand. On the other hand,
dissipation is necessary to correct the energy balance. It is this
aspect which was missing in the earlier publication \cite{ngd03}.
The external electric field provides for the energy necessary to
tunnel from an occupied site below the Fermi level to an unoccupied
site slightly above it. Generally, the gain in field energy will be
larger than the energy difference of the two levels. As dissipation
corresponds to the possibility of inelastic processes, the excess
energy can be absorbed by the dissipative bath. In the absence of an
additional bath, the conductivity
was suggested to decay in a double exponential manner with decreasing
temperature \cite{GMP04}.

The discussion of the instanton mechanism is combined with an RG
analysis of the disordered system
\cite{Glatz+Natter02,Glatz+Natter04}.  As a consequence, only
effective parameters obtained from RG enter the results.  The
combination with an RG is necessary to scale the system into a regime
where the quasiclassical instanton analysis is applicable.

We make a detailed comparison to Mott \cite{Mott69,ahl71} variable
range hopping of noninteracting particles and generalize an argument
due to Shklovskii \cite{s73} for tunneling of localized electrons in
strong fields to arbitrary dimensions.  The instanton mechanism
discussed above describes the same type of physics. We also discuss the
influence of long range interaction \cite{Efros_Shklovskii}.

We compare the physics of quantum creep to a field induced
delocalization transition discovered by Prigodin
\cite{Prigodin80,Soukoulis+83,DeSiSo84,Kirkpatrick86,PriAl89}.  This
transition is due to the energy dependence of the backscattering
probability from a single impurity.  As the energy depends linearly
on position in the presence of an external electric field, the
localization length acquires a spatial dependence.  If the
localization length grows sufficiently fast, the wave function is no
longer normalizable and the state is delocalized. We show that this
mechanism becomes effective at field strengths which are
parametrically larger than the crossover field for which quantum
creep becomes important. In addition, we discuss a generalization of
this mechanism to interacting systems.  We find that for repulsive
short range interactions the delocalizing effect of an external
field is less pronounced (in our approximation it even completely
disappears) and hence delocalization is not a relevant competition
for quantum creep. For attractive short range interactions, the
delocalizing effect of an external field is enhanced.

The organization of the paper is as follows: in section II we set up
the model and discuss its renormalization. The tunneling due to
instantons and the subsequent time evolution of the quantum sine
Gordon system is discussed as a reference point in section III.  Based
on this analysis, tunneling in the disordered case and the essential
influence of dissipation is presented in section IV. In section V we
compare with previous results for noninteracting systems obtained by
Shklovskii and Mott, in section VI we study the influence of a long
range Coulomb interaction, and in section VII we discuss the relevance
of a localization-delocalization transition in external fields to the
problem of quantum creep in 1d systems.

\setcounter{equation}{0} 
\section{Model and zero-field renormalization}
\label{sec:model_zfr}

\subsection{The model}

We consider in this paper disordered charge- or spin density waves
or Luttinger liquids with a density \cite{Gruner,Haldane,Giamarchi}
\begin{equation}\label{eq:n-zfr1}
    \rho=\left(\rho_{0}+\frac{1}{\pi}\partial_{x}\varphi\right)
    \big(1+\Delta\cos(p\varphi+Qx)\big)\,.
\end{equation}
Here $p=1,2$ for charge density waves (CDW) and Luttinger liquids
(LL), respectively, and $p\rho_{0}=(1/\pi)Q$. The form
(\ref{eq:n-zfr1}) preserves the conservation of the total charge
under an arbitrary deformation $\varphi(x)$, provided
$|\nabla\varphi|\ll Q$. The wave vector of the density modulation
$Q=2k_{F}$ for LLs and CDWs (although the value of $Q$ can be
different for CDWs depending on the precise shape of the 3D Fermi
surface and the optimal electron phonon coupling), and $Q=4k_{F}$
for spin density waves (SDWs). $\Delta=2$ for LLs with a strictly
linear dispersion relation\cite{Haldane} whereas for CDWs
$\sqrt{\rho_{0}\Delta}$ denotes the amplitude of the order
parameter of the condensate.

The Hamiltonian of the corresponding disordered system is then given by
\begin{eqnarray}\label{eq:n-zfr2}
\hat{\mathcal{H}}&=&\frac{1}{2\pi}\int dx\bigg
\{\hbar vK(\pi\hat{\Pi})^{2}+\frac{\hbar v}{K}
(\partial_{x}\hat{\varphi})^{2}\nonumber\\
&&+2\pi\rho(x)\big(-Exe_0+U_{R}(x)\big)\bigg\}\,,
\end{eqnarray}
where $\hat{\Pi}(x)$ denotes the momentum operator conjugate to
$-\hat{\varphi}(x)$:
\begin{equation}\label{eq:n-zfr3}
    \big[\hat{\Pi}(x),\hat{\varphi}(x')\big]=i\delta(x-x')\,,
\end{equation}
$v$ denotes the velocity of the phason excitations, and $K$ is the
measure of short range interactions. The quantities can be expressed
as $\frac{v}{K}=\frac{1}{\pi\hbar\kappa}$,
$vK=\frac{\pi\hbar\rho_{0}}{m}$ and hence
$K^{2}=\frac{\pi^{2}\hbar^{2}\rho_{0}\kappa}{m}$ and
$v^{2}=\frac{\rho_{0}}{m\kappa}$ where $\kappa$ is the
compressibility and $m$ the effective mass of the charge carriers
(for more details see Appendix \ref{append}).

For noninteracting spinless fermions $K=1$ and $v=v_{F}$ where
$v_{F}$ denotes the Fermi velocity. In CDWs $K$ is small, of the
order $10^{-2}$ whereas in SDWs $K$ may reach one \cite{Gruner}.
$E$ denotes the external electric field.

The random potential $U_{R}(x)$ is considered to be build of
isolated impurities at
random positions $x_{i}$, $i=1,\ldots,N_{\text{imp}}$:
\begin{equation}\label{eq:n-zfr4}
    U_{R}(x)=-u_{0}\sum\limits_{i=1}^{N_{\text{imp}}}\delta(x-x_{i})\,.
\end{equation}
In the following, we will neglect the forward scattering term
$\sim\partial_{x}\varphi U_{R}(x)$ since it does not affect the time
dependent properties. Since disorder is assumed to be weak and hence
effective  only on length scales much larger than the impurity
spacing,  we may rewrite the backward scattering term in the
continuum manner:
\begin{eqnarray}\label{eq:n-zfr5}
  \lefteqn{-\rho_0\Delta u_0\int dx
  \sum\limits_{i=1}^{N_{\text{imp}}}\delta(x-x_i)\cos(p\varphi(x)
  +Qx)}\nonumber\\
  & & \to
    -\rho_0\Delta u_0n^{1/2}_{\text{imp}}a^{-1/2}
    \int dx\cos\big(p\varphi+\alpha(x)\big)\,.
\end{eqnarray}
Here $ n_{\text{imp}}=N_{\text{imp}}/L$ is the impurity
concentration, and $\alpha(x)$ is a random phase equally distributed
in the interval $0\le\alpha<2\pi$
and $\overline{e^{i(\alpha(x)-\alpha(y))}}=a\delta(x-y)$.
For weak pinning, the
replacement (\ref{eq:n-zfr5}) is valid on length scales much larger than
the mean impurity spacing and
can be justified by the fact that both
parts of (\ref{eq:n-zfr5}) lead to the same pair correlations , e.g.
the same replica Hamiltonian
\begin{equation}\label{}
-\frac{1}{2}(\rho_0\Delta u_0)^2n_{imp}\int dx
\cos\left(p(\varphi_\alpha(x)-\varphi_\beta(x)\right).
    \end{equation}
and hence to the same physics.

 The effective
Hamiltonian can therefore be rewritten as
\begin{equation}\label{eq:n-zfr6}
\begin{split}
    \hat{\mathcal{H}}&=
    \frac{1}{2\pi}\int dx\bigg\{\hbar vK(\pi\hat{\Pi})^{2}+
    \frac{\hbar v}{K}(\partial_{x}\varphi)^{2}\\
    & -2e_0\hat{\varphi}E-
    2\pi\rho_{0}\Delta n^{1/2}_{\text{imp}}a^{-1/2}u_{0}
    \cos\big(p\varphi+\alpha(x)\big)
    \bigg\}\,.
    \end{split}
\end{equation}

Below we will use an imaginary time path integral formulation. It
is convenient to use \emph{dimensionless} space and {time}
coordinates by changing $x/a\to x$ and $v\tau/a\to y$ where $a$ is
a small scale cut-off. The Euclidean action of our system is then
given by
\begin{eqnarray}\label{eq:n-zfr7}
    \frac{S}{\hbar}&=&\frac{1}{2\pi K}\int dx\int\limits_{-K/2T}^{K/2T}dy
    \Big\{(\partial_{y}\varphi)^{2}+
    (\partial_{x}\varphi)^{2}\nonumber\\
    &&-2f\varphi-2u\cos\big(p\varphi+\alpha(x)\big)\Big\}\,.
\end{eqnarray}
The \emph{dimensionless} parameters of the theory are
\begin{equation}\label{eq:parameters}
\begin{split}
K,\,\,\,\,\,\,\,T=\frac{Ka}{\hbar\beta v},\,\,\,\,\,\,\,
f=e_0Ea\beta T=\frac{Ke_0Ea^2}{\hbar v},\\
    u=\frac{2\pi a^{3/2}K\rho_{0}\Delta n^{1/2}_{\text{imp}}u_{0}}
    {\hbar v}.
\end{split}
\end{equation}
$\beta$ is the inverse temperature. All $p$-dependent physical
quantities $Q(p;K,T,u,f)$ can be calculated from the case $p=1$ by
the relation

\begin{equation}\label{}
    Q(p;K,T,u,f)=Q(1;p^2K,p^2T,p^2u,pf).
\end{equation}

\subsection{Renormalization group analysis}

Our strategy to consider the transport properties of the present
one-dimensional system includes two steps.

(i) We first integrate out phase fluctuations on length scales
from the initial microscopic cut-off $a=1$ to $\xi=e^{l^*}$
 $1<e^{l}<\xi $
keeping $f=0$; $\xi$ is determined from the condition that the renormalized
and rescaled coefficient $u(l)$
of the nonlinear term in (\ref{eq:n-zfr7}) becomes of order of one
or larger and, typically, $K$ becomes small.

(ii) In the second step we treat the problem at nonzero $f$ in the
quasi-classical limit. Corrections  to this quasi-classical limit
arise if the renormalized $K$ is still of order one.

Since the first step is well documented in the literature we will here
quote only the results.
At zero temperature and $f=0$ the flow equations are given by
\cite{Giamarchi+Schulz,Glatz+Natter04}
\begin{equation}\label{eq:n-zfr8}
    \begin{split}
    \frac{dK}{dl}&=-\frac{1}{2}p^{4}u^{2}KB_{0}(p^{2}K)\,,\\
    B_{0}(x)&=\int\limits_{0}^{\infty}d\tau\,\tau^{2}e^{-\tau}
    \left(1+\frac{\tau^{2}}{2}\right)^{-x/4}\,.
    \end{split}
\end{equation}

\begin{equation}\label{eq:n-zfr9}
    \frac{du^{2}}{dl}=\left(3-\frac{p^{2}}{2}K\right)u^{2}\,.
\end{equation}

For completeness we also quote the flow equation in the case when
the phase $\alpha(x)$ is fixed at $\alpha=0$:
\begin{eqnarray}
 \frac{dK}{dl} &=& -\frac{1}{8\pi}p^{4}u^{2}B_{2}(p^{2}K)\,,
 \label{eq:n-zfr10}\\
 B_2(x)&=&\int\limits_0^\infty d\tau\,
dx\,
 (x^2+\tau^2)e^{-\tau}
    \left(1+\frac{\tau^{2}}{2}\right)^{-x/4}\!\!\cos x\,,
    \nonumber \\
  \frac{du}{dl} &=& \left(2-\frac{p^{2}}{4}K\right)u\,.
  \label{eq:n-zfr11}
\end{eqnarray}
Note that in this pure case the definition of $u$ in (\ref{eq:parameters})
contains an additional factor $\sqrt{n_{\text{imp}}a}$.

In both cases (\ref{eq:n-zfr8}, \ref{eq:n-zfr9}) and
(\ref{eq:n-zfr10}, \ref{eq:n-zfr11}), there is a phase
transition at $K=K_{c}(u)$ with $K_{c}(0)=6/p^{2}$ and
$K_{c}(0)=8/p^{2}$, respectively. For $K>K_{c}$ the potential
becomes irrelevant and the system is in a superconducting phase.
For $K<K_{c}$ the asymptotic $RG$-flow is to small values of $K$
and large values of $u$. For the rest of the paper we will
restrict ourselves to the case $K<K_c$.

The renormalization was done so far for zero external field.
Now we switch on the field which is a relevant perturbation and
destabilizes the
system. As in the previous section we consider  on scales
$L<\xi$ only fluctuations inside one potential valley.
The field is then rescaled according to
\begin{equation}\label{eq:resfil}
    \frac{df}{dl}=2f(l)\,.
\end{equation}

\begin{figure}[htbp]
\center\includegraphics[width=0.8\columnwidth]{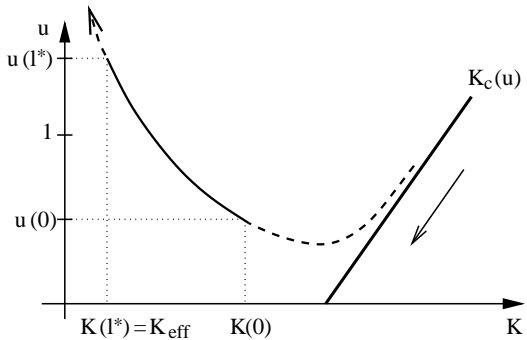}
   \caption{ Flow diagram in $K-u$ plane. RG is stopped at scale $l=l^{\ast}$
   such that $p^2u(l^{\ast})=\mathcal{M}\gg 1$.}
   \label{fig:flow}
   \end{figure}

According to our strategy we stop the flow at $l=l^{\ast}$,
$e^{l^{\ast}}=\xi$, $p^2u(l^{\ast})=\mathcal{M}$ with $\mathcal{M}\ge
1$ (see Fig. \ref{fig:flow}).  Clearly our one loop flow equations
break down as soon as $u(l)\sim 1$, but qualitatively we expect that
the flow continues to go into the direction of large $u$ and
simultaneously small $K$ such that typically $K(l^{\ast})\ll 1$.   The
restricted validity of the one-loop flow equations will then merely
result in the uncertainty of the definition of the value
$K(l^{\ast}(\cal{M}))$.


The quantities $u(l)$ and $K(l)$ are the \emph{renormalized} and
\emph{rescaled} parameters. The corresponding \emph{effective}
parameters observed on scale $L=e^l$ are unrescaled. In pure
systems $K_{\rm eff}(L)=K(l)$, $u_{\rm eff}(L)=e^{-2l}u(l)$, and
$f_{\rm eff}(L)=f=e^{-2l}f(l)$. In this way we get $u_{\rm
eff}(\xi)\approx \mathcal{M}/(p\xi)^{2}$.

For impure systems the corresponding relations are  $K_{\rm
eff}(L)=K(l)$, $u^2_{\rm eff}(L)=e^{-3l}u^2(l)$,
$f_{\rm eff}(L)=f=e^{-2l}f(l)$
and  we get $u^2_{\rm eff}(\xi)\approx\mathcal{M}/(p^2\xi^3)$.
Not too close to $K_c$, $\xi$ can be written as
\begin{equation}\label{eq:correlation_length}
    \xi\sim L_{\rm FL}^\frac{K_c}{K_c-K}
\end{equation}
where $L_{\rm FL}=a(p^2 u)^{-2/3}$
denotes the Fukuyama-Lee length. $\xi$ has the physical meaning of
a correlation length.  Note, that in the disordered case there is
no renormalization of
\begin{equation}\label{eq:v/K}
  \frac{v}{  K}=\frac{v_{\mathrm{eff}}}{ K_{\mathrm{eff}}}=
  \frac{1}{\pi\hbar\kappa}
 \end{equation}
 because of a statistical tilt symmetry \cite{USchultz}.


\setcounter{equation}{0}

\section{Tunneling -  in the pure case}\label{sec:tunneling-pc}


At low but finite temperature the lifetime $\Gamma^{-1}$ of a
metastable state is determined by quantum-mechanical tunneling
and given by the relation
\begin{equation}\label{eq:n1}
    \hbar\Gamma=-2\operatorname{Im}F=2\beta^{-1}\operatorname{Im}\ln Z\,,
\end{equation}
where $F$ is the free energy. In the
path integral formulation $Z$ is given by
\begin{equation}\label{eq:n2}
    Z=Z_{0}+iZ_{1}=\int{\mathcal D}
    \varphi (x,\tau) \,e^{-\frac{1}{\hbar}
    \int\limits_{-K_{\mathrm{eff}}/2T}^{K_{\mathrm{eff}}/2T}dy
    \,{\mathcal L}\big(\{\varphi\},f\big)}.
\end{equation}
Here ${\mathcal L}\big(\{\varphi\},f\big)$ denotes the Lagrangian
of the metastable system from which we integrated out already
fluctuations on length scales smaller than $\xi$. This resulted in
the replacement $u\to u_{\mathrm{eff}}(l^{\ast})$ and $K\to
K_{\mathrm{eff}}(l^{\ast})$. $f$ represents the driving force. The
contribution $Z_{0}$ of the partition function $Z$ contains the
contribution from field configurations restricted to stay in the vicinity
of the potential minimum. Configurations which leave the metastable
state are unstable and contribute to the imaginary part of the
free energy, which is assumed to be small such that
\begin{equation}\label{eq:n3}
    \Gamma\approx 2(\beta\hbar)^{-1}\frac{Z_{1}}{Z_{0}}\,.
\end{equation}
What we will do in the following is the calculation of the
tunneling probability in the \emph{quasi-classical} approximation:
assuming that $K_{\text{eff}}=K(l^{\ast})\ll 1$ we assume that the
only deviation from a completely classical behavior is the
generation of instantons which trigger the decay of the
metastable state. Before we come to the disordered case we
briefly exemplify the physics on the pure sine-Gordon model which
facilitates the further discussion. The imaginary time Lagrangian
of the sine-Gordon model is given by
\begin{eqnarray}\label{eq_n4}
    -{\mathcal L}\big(\{\varphi\}\big)&=
    &\int\limits_{0}^{L_0/a}dx\frac{1}{2\pi K_{\mathrm{eff}}}\bigg\{
    \left(\frac{\partial\varphi}{\partial y}\right)^{2} +
    \left(\frac{\partial\varphi}{\partial x}\right)^{2}\nonumber\\
    & &-
    2u_{\mathrm{eff}}\cos p\varphi-2f\varphi\bigg\}\,.
\end{eqnarray}
The decay of a metastable state (say $\varphi_{0}=0$) of the
sine-Gordon model has been first considered by Maki \cite{Maki}
following earlier considerations by t'Hooft \cite{tHooft} and
Coleman \cite{Coleman}. In the present framework the tunneling
process is represented by the formation of a two-dimensional
instanton which obeys the Euclidean field equation
\begin{equation}\label{eq:n5}
    \frac{\partial^{2}}{\partial y^{2}}\varphi+
    \frac{\partial^{2}}{\partial x^{2}}\varphi
    -pu_{\mathrm{eff}}\sin p\varphi+f=0
\end{equation}
and has a finite Euclidean action. The instanton solution
$\varphi_{s}(x,y)$ we are looking for forms a (spherical) droplet
in the new metastable minimum $\varphi_{1}=2\pi$. Using spherical
coordinates, $r^{2}=x^{2}+y^{2}$, we get from (\ref{eq:n5})
\begin{equation}\label{eq:n6}
    \frac{\partial^{2}}{\partial r^{2}}\varphi+
    \frac{1}{r}\frac{\partial}{\partial r}\varphi
    -pu_{\mathrm{eff}}\sin p\varphi+f=0\,.
\end{equation}
For small $f$ and hence large droplet radii $R$
 we may drop the second term. In particular for
$f=0$ the solution is
\begin{equation}\label{eq:n7}
    \varphi(r)=\frac{4}{p}\arctan\exp\big[(R-r)
    \sqrt{\pi u_{\mathrm{eff}}p^{2}}\big]
\end{equation}
which connects the original state $\varphi(r)=0$ for $r-R\gg w$
with the new (likewise metastable) state $\varphi(r)=2\pi/p$ for
$R-r\gg w$ where $w=1/\sqrt{p^2
u_{\mathrm{eff}}}\approx\xi\big/\sqrt{\cal M}$ is the width of the
droplet wall. The neglect of the second term in (\ref{eq:n6}) is
justified for $R\gg w$. In the following, we will use this narrow
wall approximation to describe the instanton only by its radius
ignoring for the moment the shape fluctuations. The action of the
instanton includes then a surface and a bulk contribution
\begin{equation}\label{eq:n8}
    \frac{S(R)}{\hbar}=\frac{2\pi}{pK_{\mathrm{eff}}}
    \left( R\sigma-R^{2}f\right)\ ,
\end{equation}
where $\sigma=\frac{8\sqrt{u_{\mathrm{eff}}}}{\pi}\approx
\frac{8}{\pi}\frac{\sqrt{\cal M}}{p\xi}$ is the surface tension of
the instanton (the Euclidean action associated with the unit
length of the instanton boundary is equal to $\hbar\sigma/(pK)$).
The action (\ref{eq:n8}) has a maximum at
\begin{equation}\label{eq:instanton_size}
R_{c}=\sigma\big/(2 f).
\end{equation}
where ${S(R_c)}/{\hbar}\equiv{S_{\text{inst}}}/{\hbar}=
{\pi\sigma^{2}}/({2pK_{\mathrm{eff}}f})$ denotes the instanton
action.

 The ratio
$iZ_{1}/Z_{0}$ appearing in the (\ref{eq:n3})
 can now be calculated
as a functional integral over circular droplets \cite{Kleinert}
\begin{equation}\label{eq:n9}
\begin{split}
    \frac{Z_{1}}{Z_{0}}\propto{\rm Im\,}
    \int \sqrt{\frac{S_{\text{inst}}}{{\hbar}}}dR\,
    e^{-S(R)/\hbar}\nonumber\\
    \propto e^{-S_{\text{inst}}/{\hbar}}
    {\rm Im\,}\int\limits_{-R_c}^{\infty}
    dr\sqrt{\frac{S_{\text{inst}}}{{\hbar}}}\,e^{2\pi
    fr^{2}/(pK_{\mathrm{eff}})}\,.
\end{split}
\end{equation}
Here we have taken into account that the change from the
integration over $\varphi$ to $R$ involves a Jacobian $\sim
\sqrt{S_{\text{inst}}/\hbar}$.
 The integral, taken along the
real axis, is divergent since the saddle point (in the functional
space) is a maximum of the action as a function of the instanton
size $R=R_c+r$.

This leads -- after the proper analytic continuation -- to the
imaginary part in $iZ_{1}$ in the partition function (for a more
detailed discussion see Schulman \cite{Schulman}). As a result we
obtain
\begin{equation}\label{eq:n10}
    \frac{iZ_{1}}{Z_{0}}\propto {i}\frac{R_c}{\xi}
    e^{-{S_{\text{inst}}}/{{\hbar}}}\,.
\end{equation}
 Further contributions to the pre-exponential
term result from the inclusion of fluctuations of the shape of the
instanton. Within the narrow wall approximation these fluctuations
could be taken into account by rewriting the Euclidean action in
the form \cite{DiehlKrollWagner}
\begin{equation}\label{eq:n13}
    \frac{S}{\hbar}=\frac{1}{pK_{\mathrm{eff}}}\sum\limits_{k=\pm}\!
    \int\limits_{-K_{\mathrm{eff}}/2T}^{K_{\mathrm{eff}}/2T}
    \!\!\!\!
    dy\left\{\sigma\sqrt{1+
    (\partial_{y}X_{k})^{2}}-2kfX_k(y)\right\}.
\end{equation}
$X_{k}(y)$ (where $k=\pm$) denote the wall position of the left and right
segment of the wall of the critical droplet, respectively. These
 fluctuations were considered  by Maki \cite{Maki}, they  lead
  to a downward renormalization of the surface tension in
(\ref{eq:n13}).

So far we considered the center of the instanton to be fixed at
$x=X_0$ and $y=0$.  However the position of the center can be
moved around which corresponds to the existence of two modes with
zero eigenvalue. Correspondingly, in the  calculation of the
partition function we have  to integrate over all possible
instanton positions which delivers in the low temperature limit
\begin{equation}\label{eq:n11}
    \frac{K_{\text{eff}}}{T}\gg R_c=\frac{\sigma}{2f}
\end{equation}
an additional factor \cite{Coleman}
\begin{equation}\label{eq:translation}
  \frac{L_0 K_{\text{eff}}}{T\xi^2}
  \frac{S_{\text{inst}}}{2\pi\hbar }\,,
\end{equation}
where $L_0$ is the sample length.
   The first factor
counts the number of different instanton positions, the second one
is the Jacobian resulting from the transition from the original to
the displacement degrees of freedom. Thus
\begin{equation}\label{eq:n12}
    \Gamma\sim
    \frac{L_0 \sigma^3 }{p\beta T f^2 \xi^3 }
    e^{-\pi \sigma^2 /(2pK_{\text{eff}}f)}\,.
\end{equation}

$\Gamma^{-1}$ represents the time scale on which the phase field
$\varphi$ tunnels  through the energy barrier between
$\varphi_{0}$ and $\varphi_{1}$.

Once the critical droplet of the new ground state $\varphi_{1}$ is
formed, the field materializes and evolves subsequently according to
the classical equation of motion \cite{Coleman}.

The latter follows from Eq. (\ref{eq:n5}) by the resubstitution
$y=ivt$. Since $\varphi_{s}(r)$ depends only on
$r^{2}=x^{2}-v^{2}t^{2}$, the solution of Eq.
(\ref{eq:n6}) also describes the evolution of the phase field
after the tunneling event. Within the narrow wall approximation,
the Minkowski action is
\begin{equation}\label{eq:n14}
    \frac{S_{M}}{\hbar}
    = -\frac{ \sigma}{pK_{\mathrm{eff}}}
    \sum\limits_{k=\pm}\int
    dt\,\bigg\{
    \sqrt{1-\frac{\dot{X}_{k}^{2}(t)}
    {v^2}}-k\frac{X_k(t)}{R_c}\bigg\}\ ,
\end{equation}
which leads to the equation of motion of two
relativistic particles
\begin{equation}\label{eq:n15}
    \pm\frac{1}{v^{2}}
    \frac{d}{dt}\frac{\dot{X}_{\pm}(t)}
    {\sqrt{1-\dot{X}_{\pm}^{2}/v^{2}}}
    =R_c^{-1}\,.
\end{equation}
This has to be solved with the initial
condition $X_{\pm}(t=0)=X_0 \pm R_{c}$
where $t=0$ corresponds to the moment of the materialization, hence
\begin{equation}\label{eq:n16}
    X_{\pm}(t)=\pm\sqrt{R_{c}^{2}+
    v_{\text{eff}}^{2}t^{2}}+X_{0}\,.
\end{equation}
Clearly, with $ivt\to y$, this is also the
saddle-point equation of action (\ref{eq:n13}), corresponding
again to a spherical droplet of radius $R_c$.

The energy ${\cal E}$ is a conserved quantity
during the motion of
the kinks as
\begin{equation}\label{eq:energy}
{\cal E} = \frac{\hbar \sigma}{p}
 \sum_{k=\pm}\Big\{\frac{1}{\sqrt{1-\dot{X}_k^2/v^2}}
 -k\frac{X_k}{R_c}\Big\}=0.
\end{equation}
Since the potential energy of the new metastable state is always
lower than that of the initial configuration the kinks will
accelerate, approaching eventually the phason velocity
$v_{\text{eff}}$.

In a long sample the nucleus of the new metastable state may form
independently at several places $X_{0}^{(n)}$ followed by a rapid
expansion of the kinks $X^{(n)}_{\pm}(t)$ which finally merge.
Below we will show that the corresponding picture in a disordered
sample is rather different.

For the sake of completeness and since we will use it later, we
consider the influence of an additional damping term in the
Euclidean action
\cite{CaldeiraLeggett,Larkin+Ovchinikov,Hida+Eckern}
\begin{equation}\label{eq:n17}
    \frac{S_{d}}{\hbar}=\frac{\eta}{4\pi}
    \int\limits_{0}^{L_0}dx
    \int\limits_{-K_{\text{eff}}/2T}^{K_{\text{eff}}/2T}
    dy\,dy'\frac{\pi^{2}\big(\varphi(x,y)-\varphi(x,y')\big)^{2}}
    {(\frac{K_{\text{eff}}}{T})^{2}
    \sin^{2}\big[\frac{T\pi(y-y')}{K_{\text{eff}}}\big]}\,.
\end{equation}
We have introduced a phenomenological dimensionless constant $\eta$
describing dissipation. This form of the Euclidean action corresponds
to a linear (Ohmic) dissipation term in the classical limit in real
time.

In the saddle point equation (\ref{eq:n5}),  (\ref{eq:n17})
results in an additional term
\begin{equation}\label{eq:n18}
    \frac{\eta}{2\pi}
    \int\limits_{-K_{\text{eff}}/2T}^{K_{\text{eff}}/2T}
    dy'\left(
    \frac{\pi T}{K_{\text{eff}}}\right)^{2}
    \frac{\big(\varphi(x,y)-\varphi(x,y')\big)^2}
    {\sin^{2}\big[\frac{T\pi(y-y')}{K_{\text{eff}}}\big]}\,.
\end{equation}
In appendix B, we justify the choice of this dissipative term and explain
its physical origin.

\setcounter{equation}{0}

\section{Tunneling - the disordered case}\label{sec:tunneling-dc}

\subsection{Surface tension of
instantons }
In this section we will consider the tunneling process in the
disordered case. In principle we follow the calculation of Section
\ref{sec:tunneling-pc}, but  important differences apply. Starting
point is again the effective action of the form (\ref{eq:n-zfr7})
on the scale $\xi=ae^{l^{\ast}}$ where $u$ is replaced by
$u(l^{\ast})$ and $K$ is replaced by $K(l^{\ast})$ as follows from
(\ref{eq:n-zfr8}) and (\ref{eq:n-zfr9}).
 The saddle point equation of
the instanton now reads
\begin{equation}\label{eq:tun-dc2}
    \left(\frac{\partial^{2}}{\partial x^{2}}+
    \frac{\partial^{2}}{\partial y^{2}}\right)\varphi
    -u(l^{\ast})\sin\big(p\varphi+
    \alpha(x)\big)+f(l^{\ast})=0\,.
\end{equation}
Let us assume that we have found its narrow wall solution  which
is parametrized by the instanton shape $X_{\pm}(y)$. Plugging this
solution  into the effective action the latter will take the
general form
\begin{equation}\label{eq_tun-dc3}
\begin{split}
    \frac{S}{\hbar}=\frac{1}{pK_{\mathrm{eff}}}\sum\limits_{k=\pm}
    \int\limits_{-K_{\mathrm{eff}}/2T}^{K_{\mathrm{eff}}/2T}dy
    \,\bigg\{
    \sigma\big(X_k,\partial_y X_{k}\big)
    \sqrt{1+(\partial_{y}X_{k})^{2}}\\
    -2kfX_{k}(y)
    \bigg\}\,.
\end{split}
\end{equation}
Here, $\sigma\big(X_k,\partial_y X_{k}\big)$ denotes the surface
tension of the instanton which depends in general both on $x$ and
on the slope $\partial_y X_{k}$ of the surface element. We could
now proceed in looking for solutions of the saddle point equation
belonging to (\ref{eq_tun-dc3}). Since this equation depends on
the specific disorder configuration a general solution seems to be
impossible. As will be shown below, the surface tension
$\sigma\big(x,0\big)\equiv \sigma_x(x)$ changes on the length
scale $\xi$ by an amount of the order $\sigma$ whereas
$\sigma\big(x,\infty\big)\equiv \sigma_y\approx\sigma$ is
constant. Changing the slope of the surface element we expect an
monotonous change of $\sigma\big(x,\partial_y X)$ between
$\sigma_x(x)$ and $\sigma_y$. Because of the rapid change of
$\sigma_x(x)$ the number of saddle point solutions will be huge.
But only one of them will be relevant for the present problem in
the sense that it rules the decay of the metastable state.


 Following previous work
 \cite{Glatz+Natter02, Glatz+Natter04},  we calculate next the
surface tensions $\sigma_x$ and $\sigma_y$ in the limit $K_{\rm
eff}\ll 1$. This is in agreement with our general strategy to take
into account quantum effects (i.e. the fact that $K_{\rm eff}\ne 0$)
only when instanton formation is considered. Additional
shape fluctuations of instantons - which are  also of quantum
origin -  could be included later on by considering the
renormalization  of this surface tension. In general they will
reduce the surface tension by a finite amount as long as we are
below $K_c$. As it was shown in \cite{Glatz+Natter02,
Glatz+Natter04} the limit $K_{\rm eff}\ll 1$ allows an
\emph{exact} solution for the (classical) ground state. To
reach this goal we rewrite the effective action
as a discrete model on a lattice with grid size $\xi$.  In the
classical ground state,  $\varphi(x,y)$ does not depend on
$y$ any more, and the renormalized form of
Eqs. (\ref{eq:n-zfr6}, \ref{eq:n-zfr7}) can be written
as
\begin{equation}\label{}
\frac{S}{\hbar}\to \frac{H_{\mathrm{eff}}}{T} \approx
\frac{1}{2\pi T}\int dx\left\{
 (\partial_{x}\varphi)^{2}-\frac{2u_{\mathrm{eff}}}{\xi^{1/2}}
    \cos\big(p\varphi+\alpha(x)\big)\right\}.
\end{equation}
Next, we replace the integration over $x$ by a summation over
discrete lattice sites $i=x/\xi$.
 This gives
\begin{equation}\label{eq:n-zfr12}
   \frac{H_{\mathrm{eff}}}{T}\approx \frac{1}{2\pi T\xi }
   \sum\limits_{i=1}^{L_0/\xi}\big\{
    (\varphi_{i}-\varphi_{i+1})^{2}-2u(l^{\ast})
    \cos(p\varphi_{i}+2\pi\alpha_{i})\big\}\,,
\end{equation}
with $0\le\alpha_{i}<1$, and $2\pi\alpha_{i}$ denoting a random
phase. If $u(l^{\ast})
=\mathcal{M}/p^2\to\infty$,
the exact ground state can be written as \cite{Glatz+Natter02,
Glatz+Natter04}
\begin{equation}\label{eq:n-zfr13}
    p\tilde{\varphi}_{i}(m)=-2\pi\alpha_{i}+2\pi
    \sum\limits_{j\le i}[\alpha_{j}-\alpha_{j-1}]_G+
    2\pi m
\end{equation}
where $[\alpha]_G$ denotes the closest integer to $\alpha$, and $m$
is an integer. A uniform electron density  corresponds to a constant
$\tilde{\varphi}_i(m)$. Disorder leads to an inhomogeneous electron
distribution with the maximum of
$\mid\tilde{\varphi}_{i+1}-\tilde{\varphi}_i\mid$ equal to  $\pi/p$
corresponding to a maximal excess charge of $\pm e_0/2$ on the scale
$\xi$.

Next, we consider a possible
\emph{bifurcation} of the ground state.
Indeed, since the $\tilde{\varphi}_i$ are constructed such that
local energy is always minimized in the ground state, we have
merely to minimize the elastic energy, which depends  on the square
of
\begin{equation}\label{}
  \frac{p}{2\pi}\big(\tilde{\varphi}_{i}(m+n)-
  \tilde{\varphi}_{i-1}(m)\big)=
  -\left(\delta \alpha_{i}-
  [\delta \alpha_{i}]_G \right)+ n,
\end{equation}
where $n$ is integer. Since  $\delta\alpha_i\equiv
\alpha_{i}-\alpha_{i-1}$ is equally distributed in the interval
$-2\le \delta\alpha_i \le 2$, we have $-{1}/{2}\le
\delta\alpha_i-[\delta\alpha_i]_G\le {1}/{2}$.  For $|\delta
\alpha_{i}-[\delta \alpha_{i}]_G|<1/2$ the ground state is uniquely
determined by $n=0$. However, for $ \delta \alpha_{i}-[\delta
\alpha_{i}]_G=k/2$, $k=\pm 1$, phase configurations with $n=0$ and
$n=k$ have the same energy and hence the ground state bifurcates.
The pairs of sites which show this property can take the values
$\tilde\varphi_{i}-\tilde\varphi_{i-1}=\mp k\pi/p$. Bifurcation
corresponds   to a local change of the charge by $\pm {2}{}e_0/p$.
Since  adding (or removing) a charge (or a pair of charges in the
case of CDWs) does not change the energy one has to conclude that
these bifurcation points correspond to states \emph{at} the Fermi
surface. The states which fulfill exactly the bifurcation condition
have measure zero, but there is a finite probability that
\mbox{$\Big||\delta\alpha_{j}|-\frac{1}{2}\Big|<\varepsilon$}. For
those pairs the cost for a deviation from the ground state is also
of the order $\varepsilon$.

\begin{figure}[htbp]
\center\includegraphics[width=0.8\columnwidth]{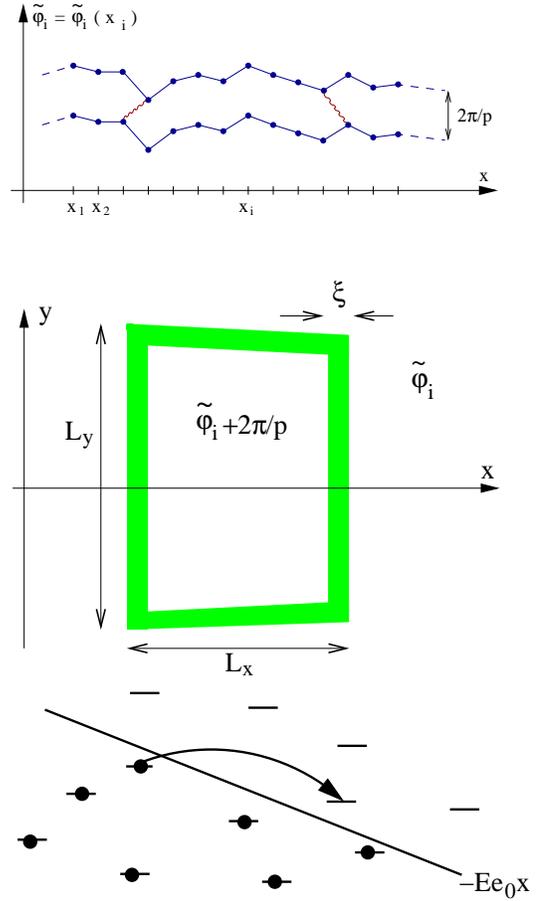}
   \caption{Upper figure: two equivalent ground states with
   two bifurcation sites which facilitate the formation of an instanton
   in nonzero external field. Middle: Top view of an almost
   rectangular instanton, the width of the wall corresponds to
   $\xi$, i.e. the distance between the grid points of the upper
   figure. Lower figure: Hopping process corresponding to the
   instanton formation, see also Fig.3 in  \cite{s73}.}
   \label{fig:shklovskii}
   \end{figure}

To calculate the surface tension
we first consider
a wall parallel to the $y$-axis.  Such a wall -
at which $m$ is changed to $m+n$, $n=\pm 1$ - has an excess energy
\begin{equation}\label{surface_tension_discrete}
\begin{split}
  \frac{H_{\text{kink}}}{T}\approx
\frac{1 }{2\pi\xi T}\Big\{\big(\tilde{\varphi}_{i}(m+n)-
  \tilde{\varphi}_{i-1}(m)\big)^2-\\
  -\big(\tilde{\varphi}_{i}(m)-
  \tilde{\varphi}_{i-1}(m)\big)^2\Big\}
=\\
  =\frac{2\pi }{p^2\xi
  T}\Big\{1-2n\big(\delta\alpha_i-[\delta\alpha_i]_G\big)\Big\}
  \equiv\frac{1}{pT}\sigma_{x,i}(n)
\end{split}
\end{equation}
On the rhs we introduced  the surface tension of the \emph{oriented}
($n=\pm1$) wall $\sigma_{x,i}(n)\equiv \sigma \nu_i(n)$, $
\sigma\approx {2\pi }/({p\, \xi})$ and \cite{ngd03}
\begin{equation}\label{eq:nu}
  \nu_i(n)=1-2n\big(\delta\alpha_i-[\delta\alpha_i]_G\big)\,,
  \,\,\,\ \nu_i(1)+\nu_i(-1)=2
\end{equation}
and hence $0\le \nu_i(n) \le 2$. 
Thus we find that  if we proceed in $x$-direction the surface
tension $\sigma_{x,i}(n)$ of a wall with a fixed orientation is
fluctuating from site to site and equally distributed in the
interval $0\le \sigma_{x,i}(n) \le 2\sigma$. If the surface tension
for a $+$-wall is close to $2\sigma$ the corresponding surface
tension for the minus wall is close to zero. If we consider the
surface tension of an oriented wall in a \emph{typical} region (i.e.
we exclude for the moment regions where rare events take place) of
linear extension $L_x\gg \xi$, then the average m-th lowest value of
$\nu_{\pm}(x)$ in this region is given by $2m \frac{\xi}{L_x}$.

An analogous calculation can be done for the
surface tension $\sigma_{y}$
if we introduce also a lattice of grid size $\xi$ in the
$y$-direction. Since the disorder is frozen in time,
$\delta\alpha_i\equiv 0$ and hence $\sigma_{y}(n)=\sigma$. Note
that the value of $\sigma$ found here agrees up to a numerical
constant with the surface tension obtained for the pure case in
the previous section.

\subsection{Instanton action in the disordered case}

To describe the
decay of the metastable state
- which is in the present case one of the classical ground states
$\tilde\varphi_i(m)$  equation (\ref{eq:n-zfr13}) - the functional
integration in $Z_1$ has to include the integration over  droplet
configurations of the new phase $\tilde\varphi_i(m+1)$ starting
with small droplets which subsequently enlarge in both  $x$ and
$y$-direction until the new metastable state is reached at least
\emph{in a part} of the $1d$ system. Since the surface tension
$\sigma_x$ can  become arbitrarily low it is obvious that the
dominating droplet configurations are those  bounded by walls
essentially parallel to the $y$-axis. Once the saddle point (the
instanton) is reached, the droplet will expand only in
$y$-direction since further expansion in $x$-direction is
prohibited by the high cost of leaving the wall position with
extra low surface tension. The maximal extension of the droplet in
$x$-direction will be denoted by $L_{x,c}$.

We will start our quantitative consideration by restricting
ourselves to \emph{rectangular} droplets of linear extension $L_x$
and $ L_y$, respectively. The instanton action $S(L_x, L_y)$ is
then given by
\begin{equation}\label{eq:tun-dc5}
\begin{split}
    \frac{S(L_x, L_y)}{\hbar}
    \frac{pK_{\text{eff}}}{\sigma}\equiv \tilde S(L_x,
    L_y)=
    L_{x}\big\{2+\eta \ln \frac{L_y}{\xi}\big\}+\\
    +\big\{\nu_+(x)+\nu_-(x+L_x)-
    R_c^{-1}L_{x}\big\}L_{y}\ ,
\end{split}
\end{equation}
where we have included also the contribution of the dissipation
\cite{Hida+Eckern}. We will assume here that the dissipation is
weak, $\eta \ll 1$. It should be noted  that the dissipation
favors in general a rectangular shape of the instanton even in the
absence of disorder \cite{Hida+Eckern}. $\nu_{\pm} (x)$ denotes
the generalization of $\nu_i(n=\pm 1)$ to a continuum description
and is a random quantity equally distributed in the interval $0\le
\nu_{\pm}\le 2$ with a correlation length of the order $\xi$.
$R_c$ is still given by (\ref{eq:instanton_size}). The functional
integral corresponding to (\ref{eq:n9}) takes now the form
\begin{equation}\label{eq:tun-dc6}
    \frac{Z_{1}}{Z_{0}}\sim
    \operatorname{Im}\int dL_{x}dL_{\tau}\,e^{-S(L_{x},L_{\tau})/\hbar}\,.
\end{equation}

\medskip
 It is illustrative to consider the
evaluation of (\ref{eq:tun-dc6}) for the \emph{pure} case, where
$\nu_{\pm}(x)\equiv 1$, and $\eta\equiv 0$, ignoring for the
moment the fact that the consideration of circular droplets is
here more natural.  The main contribution to (\ref{eq:tun-dc6})
comes now from quadratic droplets since for a given area the
square has the least circumference. At the saddle point
$L_{x,c}=L_{y,c}=2R_c$ and the instanton action is $\tilde S_c =4
R_c$. Clearly, circular droplets  have a lower saddle point action
$\tilde S_c = \pi R_c$.

 In the \emph{impure} case the situation is different because of
the variation of the surface tension $\nu_{\pm}(x)$. The saddle
point will be determined by  low values of the surface tension
$\nu_+(x)+\nu_-(x+L_{x})
$.  Since the surface tension $\nu(x)$ fluctuates on the scale
$\xi$, the droplet will not - as in the pure case - expand  both in
the $x$- and in the $y$-direction. For the further discussion it is
useful to use the parametrization $L_x=\Lambda \cdot l_x$ where $0<
l_x \le 1$. In an $x$-interval of extension $\Lambda$ the typical
surface tension fulfills  the inequality
\begin{equation}\label{eq:surface_tension}
    2\frac{\xi}{\Lambda}\lesssim 
    \nu_+(x)+\nu_-(x+L_{x})
    \equiv 2\frac{\xi}{L_x} \mu( x,l_x)\lesssim 2.
\end{equation}
For a \emph{typical} region $\mu(x,l_x)\approx  \mu( l_x)$
will not depend on the choice of $x$. 
 As explained above $\mu(l_x)$ is strongly fluctuating with a
correlation length $\xi/\Lambda$.
Since we have no detailed information about the strongly fluctuating
surface tension $\mu(l_x)$ we now minimize $S$, Eq.
(\ref{eq:tun-dc5}),  with respect to the scale factor $\Lambda$ and
$L_y$, and choosing  such a value of $l_x$ that $\mu(l_x)$ is of the
order one. This gives the saddle point equations:
\begin{equation}\label{eq:saddlepointequation}
\begin{split}
L_x(2+\eta\ln\frac{L_y}{\xi})-(2\frac{\xi  }{L_x}
\mu(l_x)+\frac{L_x}{R_c})L_y=0,\\
\eta\frac{L_x}{L_y}+2\frac{\xi}{L_x}\mu(l_x)-\frac{L_x}{R_c}=0.
\end{split}
\end{equation}
Note, that the second equation describes the energy conservation
during tunneling.  These equations  have the solution
\begin{equation}\label{eq:saddlepointsolution}
\begin{split}
L_{x,c}^2=2\xi R_c\mu(l_x)\frac{1+\frac{\eta}{2}
\ln(\frac{eR_{c}}{\xi})}{1+\frac{\eta}{2}\ln(\frac{R_{c}}{e\xi})},\\
L_{y,c}=R_c\left(1+\frac{\eta}{2}\ln\left(\frac{eR_{c}}{\xi}\right)\right).
\end{split}
\end{equation}
Thus we get for the saddle point action
\begin{equation}\label{eq:saddlepointaction}
  \frac{S_c}{\hbar}=\frac{2\sigma}{pK_{\text{eff}}}
  \sqrt{\mu(l_x)\xi R_c}\sqrt{\left(1+\frac{\eta}{2}
  \ln\frac{R_{c}}{\xi}\right)^2-\left(\frac{\eta}{2}\right)^2}.
\end{equation}
This solution becomes exact in the quasiclassical limit $K_{\rm eff} \ll 1$.
For $K_{\rm eff} \lesssim 1$, there are additional fluctuations which lead to
downward renormalization of the surface tension.
\subsection{Stability of the saddle point solution}
\medskip
Considering quadratic fluctuations around the saddle point it is
easy to show that one eigenvalue is negative as it has to be the
case for an instanton. The eigenvector corresponding to this
eigenvalue has a small component in $x$-direction which could
suggest a further growth of the droplet both in $y$- \emph{and} in
$x$-direction. For the further discussion it is however important to
realize that the decay from the true saddle point has to follow
\emph{one} of the many surface tension minima inscribed in the
function $\mu(l_x)$ and which determines the saddle point value of
$l_x$. The latter would follow from a more microscopic approach
using a variation with respect to $L_x=\Lambda l_x$. Such a
treatment would require the detailed knowledge of the energy
landscape $\mu(l_x)$ which is unknown. Since $\mu(l_x)$ is of the
order of one for the low surface tension valleys we conclude the for
the true saddle point $l_{x,c}\equiv l_c$ and $\mu(l_c)\equiv \mu_c
$ remain unspecified, but are both of order unity. Such a decay
within one energy valley can only happen if (i) the slope of $S$ in
$y$-direction is negative and (ii) if
\begin{equation}\label{eq:decay}
   \frac{ \partial^2S}{\partial L_y^2}|_{L_y=L_{y,c}}
   =-\eta \frac{L_x}{L_y^2}<0.
\end{equation}
The first condition is fulfilled, as one can see from the saddle
point equation for $L_y$. The second condition depends crucially on
the existence of dissipation. Without dissipation, e.g. the emission
of phonons, the decay to the new metastable state is impossible
because it would violate energy conservation. In experimental
systems, dissipation is always present, for example due to electron
phonon coupling. The precise way in which the dissipation strength
enters the final result as a prefactor is beyond the logarithmic
accuracy of our present calculation. However, from a comparison with
the decay rate of dissipative two level systems \cite{Weiss99} we
conjecture that the decay rate is proportional to the damping
strength $\eta$. The answer to the question whether an short range
interacting and disordered 1D electron system has a sufficient
amount of intrinsic dissipation to support a finite creep current is
clearly beyond the scope of the present work.

As a final result we get for the decay rate $\Gamma(E)$ in the limit
of $\eta\ll 1$, $E\ll E_0$ and hence for the nonlinear zero
temperature conductivity $\sigma(E)$
\begin{equation}\label{eq:decay_rate}
 \ln \Gamma(E) \sim \ln\sigma(E) \sim -
2\frac{L_x(E)}{\xi_{\text{loc}}}
\left(1+{\eta}{}\ln\frac{L_x(E)}{\xi}\right)
\end{equation}
where
\begin{equation}\label{eq:decay_rate2}
 \frac{L_x(E)}{\xi_{\text{loc}}}\approx
\sqrt{\frac{E_0}{E}},\,\,\, \,\,\,\,\,E_0=\frac{\pi\hbar
v}{pKe_0\xi_{\text{loc}}^2}=\big(p\kappa
e_0\xi_{\text{loc}}^2\big)^{-1}\,\,\,\,.
\end{equation}
$L_x(E)$ denotes the extension of the instanton and hence the
distance over which charges are tunneling in $x$-direction.
$\kappa$ is the compressibility which depends on the interaction
parameter $K$ (compare Appendix A). $\xi_{\text{loc}}=p^2\xi
K_{\text{eff}}$ is the localization length of the tunneling
charges,   in agreement with a result of Fogler \cite{fogler02}.
In fact the correlation length $\xi$ was used in
\cite{Giamarchi+Schulz,ngd03} and subsequent papers as the
\emph{localization} \emph{length}, which is correct for
intermediate $K$. In the limit $K\to 0$, $\xi$ goes over into the
Fukuyama-Lee length which is a completely classical quantity.
Tunneling processes are characterized by a length scale
$\xi_{\text{loc}}$ which is intrinsically  of quantum mechanical
origin and  therefore has  to vanish for $\hbar\to 0$.

\subsection{Percolating instantons}


\medskip
Once a single instanton was formed it will expand in the
$y$-direction leaving a region of linear extension $L_x(E)$ behind
in which the phase is advanced by $2\pi/p$. Since such a
configuration corresponds to a change of the phase
\emph{difference} of neighboring sites at the left and right
boundary of this region by $\pm 2\pi/p$ one has to conclude that
such an event corresponds to a transfer of a charge $2e_0/p$ from
one boundary to the  other. Many such events will happen
independently at different places of the sample, each of it leads
to a local transfers of charge over a typical distance $L_x(E)$. A
current will only flow if these tunneling processes fill
eventually the whole sample, see Fig.~3.

\begin{figure}[htbp]
\center\includegraphics[width=0.95\columnwidth,height=0.4\columnwidth]
{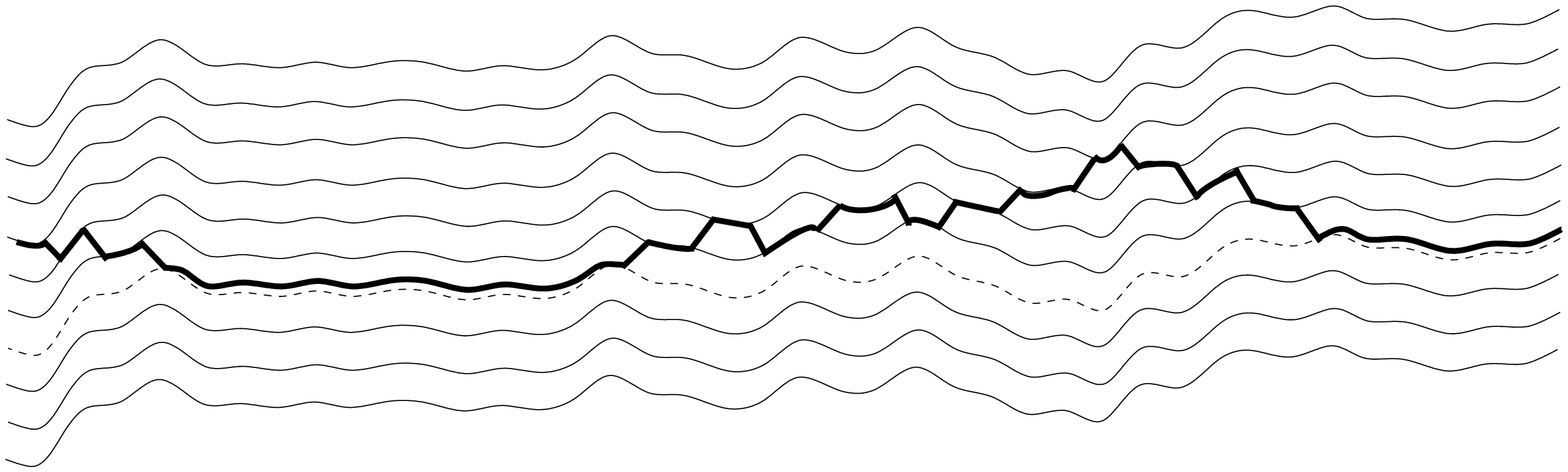}
   \caption{Phase profile $\varphi(x)$ (bold line) after
   several tunneling events have happened, starting from
    the zero field ground state characterized by the dashed line.
   The thin lines show different zero field ground states
   which follow from each other by phase shifts of $2\pi/p$. A
   current flows only if the instantons form a percolating path
   through the sample.
   }
   \label{fig:qeq}
   \end{figure}

It is in this respect important to note that bifurcation sites (at
which the original metastable state $\tilde\varphi_{j}(m),\forall
j$ was changed to $\varphi_{j\ge i}=\tilde\varphi_j(m+n)$
($n=\pm1$)) are now favorable sites for a change to
$\varphi_{j}=\tilde\varphi_j(m+n), \forall j$. Indeed, if the
initial reduced surface tension for this pair of sites was
(compare(\ref{surface_tension_discrete}))
$\nu_i=
\epsilon$, then we get for the surface tension between the state
$\tilde\varphi_{j}(m+n), \forall j$ and $\varphi_{j\ge
i}=\tilde\varphi_j(m+n), \varphi_{j<i}=\tilde\varphi_j(m)$

\begin{equation}\label{surface_tension_discrete_2}
\begin{split}
  \frac{H_{\text{kink}}}{T}\approx
\frac{1}{2\pi\xi T}\Big\{\big(\tilde{\varphi}_{i}(m+n)-
  \tilde{\varphi}_{i-1}(m+n)\big)^2-\\
  -\big(\tilde{\varphi}_{i}(m+n)-
  \tilde{\varphi}_{i-1}(m)\big)^2\Big\}=\\
  =\frac{2\pi}{p^2\xi T}
  \Big\{-1+2n\big(\delta\alpha_i-[\delta\alpha_i]_G\big)\Big\}.
\end{split}
\end{equation}
The reduced surface tension of the new wall is therefore
\begin{equation}\label{eq:nu1}
  \nu_i(n)=-1+2n\big(\delta\alpha_i-[\delta\alpha_i]_G\big)= -\varepsilon\ ,
\end{equation}
i.e. these sites are preferential for the formation of new
instantons which will fill the gaps between already existing
regions of increased $\varphi$. These considerations have a simple
interpretation in terms of the  charges which undergo tunneling:
In the initial ground state $\tilde\varphi_i(m)$  all occupied
states have energies below the Fermi level. The creation of an
instanton corresponds to the transfer of a charge ($e_0$ for
Luttinger liquids and $2e_0$ for CDWs) from an occupied to an
unoccupied site of distance $L_x(E)$. The site which is unoccupied
now is below the Fermi-energy and consequently its surface tension
is negative.

\medskip

As one can see from (\ref{eq:decay_rate}) the current $I\sim
 -\frac{1}{\pi}\langle\dot{\varphi}\rangle\sim\Gamma(E)$ shows for $E\ll E_0$
 creep-like behavior. The calculation of the complete pre-factor of
 the decay rate $\Gamma$ is beyond the scope of the present paper.


\subsection{Finite temperatures}

So far we assumed that the typical extension $L_{y,c}$ of the
instanton in $y$-direction is much smaller than
$K_{\text{eff}}/T$. However, this condition is violated for
\begin{equation}\label{eq:thermal_crossover}
    \beta<\beta_E\equiv \frac{p}{e_0E\xi_{\text{loc}}}\,\,\,\,
    \text{or}\,\,\,\, E<E_{\beta}=\frac{p}{e_0\xi_{\text{loc}}\beta}.
\end{equation}
For higher temperatures, $\beta<\beta_E$, the saddle point cannot
be reached any more. In the limit $\beta\ll \beta_E$ we can
proceed as in \cite{ngd03} and consider the contribution to the
average current $j\sim -\frac{1}{\pi}\langle\dot{\varphi}\rangle$
from the production of droplets of the metastable states with
$n=\pm 1$.This leads in the limit
$\beta e_0 EL_x\ll 1$
to a linear relation
between $j$ and $E$. The maximal barrier is now given by
$L_y\approx K_{\text{eff}}/T$, $L_x\approx L_x(E_{\beta})$ which
gives a \emph{linear} conductivity
\begin{equation}\label{eq:linear_conductivity}
\sigma(T)\sim \exp\big\{-\text{const.}\sqrt{\hbar
\omega_p\beta}/K_{\text{eff}}\big\}\,.
\end{equation}
Here,  $T = \pi \kappa a/\beta$ where $\beta$ is the inverse
temperature, and we introduced the pinning frequency $\omega_p$ by
\begin{equation}\label{eq:pinning_frequency}
\omega_p=\frac{\pi p v_{\text{eff}}}{\xi}.
\end{equation}
The result (\ref{eq:linear_conductivity}) can be also written in
the form
\begin{equation}\label{eq:linear_conductivity_2}
\sigma(T)\sim \exp\big\{-\text{const.}\sqrt{\beta
/(\kappa\xi_{\text{loc}})}\big\}.
\end{equation}
where we used the compressibility $\kappa$ defined in the Appendix.
Note that $\kappa$ depends on the interaction parameter $K$.
Equation (\ref{eq:linear_conductivity_2}) is the natural extension
of the Mott variable range hopping result to short range interacting
electrons. A long range Coulomb interaction leads to a
logarithmic correction in Eq.~(\ref{eq:linear_conductivity_2}),
see \cite{teber_fogler_shklovskii}. The additional logarithmic correction
due to a dissipative damping  of the tunneling action by electronic degrees
of freedom \cite{teber_fogler_shklovskii}
 is only relevant for strong pinning, whereas in the situation
of weak pinning this correction is absent as all electronic degrees of
freedom are localized and cannot dissipate energy.


\subsection{Rare events}

So far we considered typical instantons. The question arises what is
the influence of regions in which the lowest surface tension is
untypically large? The question was addressed for nonzero temperatures
\cite{Kurkijarvi,SerotaKaliaLee,RaikhRuzin}. We proceed in a similar manner.
The regions with large surface tension
will lead to a lower tunneling probability and hence act as
a weak link.  We may first ask the question: what is the
\emph{largest} \emph{distance} $x_m$ between two consecutive sites
of surface tension $\nu_i\le\nu_0$ in a sample of length $L_0$?
These sites are randomly distributed along the sample of length
$L_0$. Their concentration is given by
\begin{eqnarray}
n_0 = \nu_0/\xi\,. \label{nu0density}
\end{eqnarray}
The probability distribution 
for the separation $x$ between
such consecutive sites is
\begin{eqnarray}
p_0(x)=n_0 e^{-n_0 x}\,. \label{distrdensity}
\end{eqnarray}
The number of sites with $\nu_i<\nu_0$ has the average $L_0n_0$ and
a standard deviation $\sqrt{L_0n_0}$. Therefore, the largest
separation $L_{x,m}$ between such sites can be found from
\begin{eqnarray}
L_0n_0 \int\limits_{L_{x,m}}^\infty dx\, p_0(x)\sim 1\,,
\end{eqnarray}
where the integral represents the probability that the given pair of
neighboring sites has the separation exceeding $L_{x,m}$.
  Using Eqs. (\ref{nu0density}) and (\ref{distrdensity}), we find:
\begin{eqnarray}
\nu_0=\frac{\xi}{L_{x,m}}\ln\left(L_0
\frac{\nu_0}{\xi}\right)\approx \frac{\xi}{L_{x,m}}\ln\left(
\frac{L_0}{L_{x,m}}\right)\,. \label{xm}
\end{eqnarray}
Comparing this equation with (\ref{eq:surface_tension}) we see that
in the region considered $\mu(x,l_x)$ is of the order $\ln(L_0
/L_{x,m})$ instead of order one as in the typical regions. Combining
this result with the saddle point equations
(\ref{eq:saddlepointequation}) we find that we have use the
replacement
\begin{equation}\label{}
\mu(l_x) \rightarrow \mu(l_x)\ln(L_0 /\sqrt{\xi R_c})
\end{equation}
in (\ref{eq:saddlepointsolution}) and the subsequent formulas.
Since it is this weak link which will control the total current ,the
main exponential dependence of the current is given by
\begin{eqnarray}\label{current-rare_events}
\ln I \sim -2
\sqrt{\frac{E_0}{E}}\sqrt{\ln\left(\frac{L_0}{\xi_{\text{loc}}}
{\sqrt{\frac{E}{E_0}}}\right)} \,.
\end{eqnarray}

We may now consider the influence  of this weakest link on the
conductivity at \emph{finite} temperatures. We found before that the
nonlinear conductivity crosses over to a linear behavior when the
extension of the instanton in $y$-direction reaches
$K_{eff}/T=L_{y,c}\approx R_c$. This relation is not affected by the value
of the surface tension as can be seen from
(\ref{eq:saddlepointsolution}). We have therefore to use the same
replacement $E\rightarrow E_{\beta}$ as for the typical instantons
but take into account the extra factor in
(\ref{current-rare_events}), which gives for the resistance of the
the sample of length $L_0$
\begin{eqnarray}
\ln(R_m/R_0) \sim \sqrt{\frac{\beta}{\kappa\xi_{\rm loc}}}
\sqrt{\ln\left(L_0\sqrt{\frac{\kappa}{\beta\xi_{\rm loc}}}\right)}
\,.
\end{eqnarray}

Finally we mention that for exponentially large samples $L_0\sim\xi
e^{R_c/\xi}$ there are weak links with surface tension $\nu_0$ of
the order one. In this case the tunneling probability is $\ln
\Gamma(E)\sim E_0/E$. A similar crossover with respect to the sample
length was found in Refs. \onlinecite{SerotaKaliaLee,RaikhRuzin}
for linear conductivity in 1d in the case of nonzero temperature.


  \section{Shklovskii and Mott Variable range hopping}
  \label{sec:VRH}
\setcounter{equation}{0}

In this Section we give a brief account of the derivation of
Shklovkii's zero temperature nonlinear hopping conductivity which
was done for three dimensions \cite{s73}. Its extension to general
dimensions $d$ is straightforward.  All electrons are assumed to
be localized over a distance of the order of the localization
length $\xi_{\text{loc}}$. An electron can be transferred from a
filled to an empty site separated by  a distance $L_{x}$ without
absorption of a phonon provided the difference between the
difference of the energies of the states does not exceed $e_0 E
L_{x}$. If we denote the density of states (per energy and unit
length) by $g(\cal{E})$, the number of states states with energy
smaller than $e_0 E L_x$ accessible  by a jump over the spatial
distance $L_x$ is given by $g({\cal E}_F)\cdot e_0 E L_{x}^{d+1}
\gtrsim 1$. $\cal{E}_F$ denotes the Fermi energy. The minimal
distance for the tunneling is therefore given $L_{x,c}\approx
\big(e_0 E g({\cal E}_F) \big)^{-1/(d+1)}$. The probability for an
electron to jump over a distance $L_{x,c}$ is $P(L_x)\sim
e^{-2L_{x,c}/\xi_{\text{loc}}}\sim I$ and hence we get for the
nonlinear current voltage relation
\begin{equation}\label{eq:Shklovskii_VRH}
    I(E)\sim e^{-\big[{E_0}/{E}\big]^{1/(d+1)}},\,\,\,
    E_0=\left(\frac{2}{\xi_{\text{loc}}}\right)^{d+1}
    (e_0g({\cal E}_F))^{-1} \ .
\end{equation}

This result agrees  with our findings (\ref{eq:decay_rate}) in
$d=1$ dimension in the noninteracting case $K=1$.

Indeed, the one-particle density of states at the Fermi level
can be written as
\begin{equation}\label{eq:DOS}
g({\cal E}_F)=\frac{1}{L_0}\sum\limits_k \delta({\cal E}_F-\epsilon_k)\,,
\end{equation}
where $\epsilon_k$ are the energies of the eigenstates of the Hamiltonian
describing noninteracting electrons, and $L_0$ is the system size.
At zero temperature, the particle density $\rho$ can be expressed as
\begin{equation}\label{eq:dens}
\rho(\mu)=\frac{1}{L_0}\sum\limits_k \Theta({\cal E}_F-\epsilon_k)\,,
\end{equation}
and hence $\kappa\equiv \partial\rho/\partial\mu=g(\mu)$
for $\mu={\cal E}_F$.

In the case of finite temperatures when
$e_0E\xi_{\text{loc}}\beta<1$ the number of reachable sites is
determined by thermal activation \cite{Mott69,ahl71}. Indeed, the number
of these sites located in a volume $L_x^d$ and the  energy
interval $\Delta E$ is given by $g({\cal E}_F)\Delta E L_{x}^{d}
\gtrsim 1$. The hopping probability is then to be determined from
the maximum of
\begin{equation}\label{}
P(L_x)\sim
e^{-2L_{x,c}/\xi_{\text{loc}}}e^{-\beta/(g({\cal E}_F)L_{x}^{d})}\ ,
\end{equation}
which gives
\begin{equation}\label{}
L_{x,\beta}=
\Big(\frac{\beta d\xi_{\text{loc}}}{2g({\cal E}_F)}\Big)^{1/(d+1)}\,,
\end{equation}
and hence for the current
\begin{equation}\label{eq:Mott_VRH}
    I\sim E\cdot \exp\Big[{-\frac{2(d+1)}{d}
    \Big(\frac{\beta
    d}{2g({\cal E}_F)\xi_{\text{loc}}^d}\Big)^{1/(d+1)}}\Big]\,.
\end{equation}

Thus there is a crossover from a nonlinear current voltage
relation at low temperatures
$\beta>\beta_E\equiv(e_0E\xi_{\text{loc}})^{-1}$ to a linear relation
at higher temperatures. The result (\ref{eq:Mott_VRH}) agrees for
$K=1$ again with our result (\ref{eq:linear_conductivity}).


\setcounter{equation}{0}
\section{The influence of Coulomb interaction}


Efros and Shklovskii \cite{Efros_Shklovskii} considered the
influence of the Coulomb interaction on Mott variable range hopping.
The Coulomb repulsion leads in dimensions $d>1$ to a suppression of
the density of states close to the Fermi energy, $g({\cal E})\sim
|{\cal E-E}_F|^{d-1}$, which changes the exponent $1/(d+1)$ in
(\ref{eq:Shklovskii_VRH}) and (\ref{eq:Mott_VRH}) to $1/2$ in all
dimensions. This can be seen as follows:

To calculate the density of states in a system with Coulomb
interaction we first consider deviations from the ground state.
For any pair of localized states close to the Fermi surface the
transfer of an electron from am occupied site ${\bf r}_{i}$ of
energy ${\cal E}_i <{\cal E}_F$  to an empty site ${\bf r}_{j}$ of
energy ${\cal E}_j
>{\cal E}_F$ the net change of energy has to be positive
\begin{equation}\label{energy_balance}
\Delta{\cal E}={\cal E}_j-{\cal E}_i -\frac{e_0^2}{\epsilon_s
\mid{\bf r}_{i}-{\bf r}_{j}\mid}>0.
\end{equation}
Here $\epsilon_s$ denotes the dielectric constant. Let us now
assume that we consider all energy levels with $\mid{\cal
E}_i-{\cal E}_F\mid<{\tilde{\cal E}}$. Apparently,  they have to
fulfill the inequality (\ref{energy_balance}) which gives
\begin{equation}\label{energy_balance1}
 \mid{\bf r}_{i}-{\bf r}_{j}\mid \gtrsim\frac{e_0^2}{\epsilon_s\tilde{\cal E}}.
\end{equation}
Hence  the donor concentration $n({\tilde{\cal E}})$ cannot be
larger than $n({\tilde{\cal E}})\lesssim
\big(\frac{e_0^2}{\epsilon_s\tilde{\cal E}}\big)^{-d}$, from which
we find for the density of states $g(\tilde{\cal
E})=\frac{dn(\tilde{\cal E})}{d{\tilde{\cal E}}}\approx {\tilde{\cal
E}}^{d-1}\epsilon_s^d/e_0^{2d}$. Thus there is no change in the
density of states in $d=1$ dimensions. A slightly more refined
calculation gives a logarithmic correction to the density of states
which give logarithmic modifications of in the exponents of
(\ref{eq:Shklovskii_VRH}) and
(\ref{eq:Mott_VRH})\cite{teber_fogler_shklovskii,teber}.

We consider now the influence of \emph{long} \emph{range} Coulomb
interaction. 
In the Fourier transformed action (\ref{eq:n-zfr7}) $
k^2|\varphi_k|^2$ is replaced by $ k^2(1+{\cal
C}\ln(k_F/k))|\varphi_k|^2$ where the dimensionless prefactor ${\cal
C}=\frac{2\pi e_0^2K}{\hbar v\epsilon_s}= \frac{2\pi^2
e^2_0\kappa}{\epsilon_s}$ measures the relative strength of the
Coulomb interaction. Fluctuations of $\varphi$ are now reduced
leading to a  suppression of the transition to the delocalized
superconducting phase. The effect of the Coulomb interaction in the
localized phase is weak and we have still a flow of our coupling
parameter $u$ to large values. Following our considerations of
Section IV we have next to construct the exact ground state in the
presence of Coulomb interaction. Because of its nonlocal character,
Eq. (\ref{eq:n-zfr13}) is not longer the true ground state. However,
it is clear that any configuration which deviates from the ground
state has to increase the energy. We consider now the instanton as
such a configuration. The effective instanton action is now given
by
\begin{equation}\label{eq:action_with_Coulomb}
\begin{split}
    \frac{S(L_x, L_y)}{\hbar}
    \frac{pK_{\text{eff}}}{\sigma}
    =
    L_{x}\big\{2+\eta \ln \frac{L_y}{\xi}\big\}+\\
    +\big\{\big({\mu}_{}(x, l_x)-{\tilde{\cal
    C}\big)\frac{\xi}{L_x}}-
    \frac{L_{x}}{R_c}\big\}L_{y}.
\end{split}
\end{equation}
Here ${\mu}_{}$ denotes now the surface tension in the presence of
the Coulomb interaction and there is an additional contribution
$\sim \tilde{\cal C} \sim {\cal C} $ from the Coulomb interaction
between the surfaces. Since we consider deviations from the ground
state, ${\mu}_{}(x, l_x)-{\tilde{\cal C}}$ has to be positive. It is
therefore tempting to assume that this difference scales as
${\mu}_{}(x,l_x)-{\tilde{\cal C}}\approx {\mu}_{\cal C}( l_x)$ and
the results of Section IV apply with ${\mu}( l_x)$ replaced by
${\mu}_{\cal C}( l_x)$, in complete agreement with the consideration
for the case without long-range interactions. Thus, the Coulomb
interaction does not change the power law dependence of the the
results (\ref{eq:decay_rate}) and (\ref{eq:linear_conductivity_2})
on $E$ and $\beta$.
There is a  logarithmic corrections to $\mu_{\cal C}( l_x)$ due
to  long range Coulomb interactions \cite{teber}
 which is beyond the accuracy of the present
calculation.


\setcounter{equation}{0}
\section{Delocalization in strong fields}\label{sec:delocalization}

For our discussion of quantum creep in 1d electronic systems the
presence of dissipation was essential. A possible nonlinear I-V
characteristic in noninteracting electron systems \emph{without}
dissipation was discussed by Prigodin \cite{Prigodin80} and others
\cite{Soukoulis+83,DeSiSo84,Kirkpatrick86,PriAl89}.  This
nonlinearity is due to a field dependent delocalization of
electronic wave functions. To understand whether this mechanism
possibly interferes with the quantum creep discussed in section IV.,
we briefly review its derivation and discuss a generalization to
short range interacting electron systems afterwards.

The disordered potential (2.4) consists of a series of delta
function scatterers of strength $u_0$ placed at irregular
positions $x_i$ with a mean density $n_{\rm imp}$. A free electron
incident  with energy $\cal{E}$ on one of these scatterers is
reflected back with a probability
%
\begin{equation}
R({\cal E}) = \Big[ 1 + \frac{2 {\cal E} \hbar^2} {m
u_0^2}\Big]^{-1} \ .
\end{equation}
%
Neglecting localization effects for the moment, the energy
dependent mean free path is given by $\ell ({\cal E}) = {
R^{-1}({\cal E})}  {n^{-1}_{\rm imp}} $. In the presence of an
external electric field, the electron kinetic energy becomes
position dependent according to ${\cal E}(x) = {\cal E}_0 + e E
x$, leading to a position dependent mean free path
%
\begin{equation}
\ell (x) = \left[ 1 + \frac{ 2 \hbar^2}{m u_0^2} ( {\cal E}_0 +
e_0 E x) \right] \frac{1}{n_{\rm imp}} \ \  \label{meanfree.eq}
\end{equation}
%
for noninteracting electrons.
 In one-dimensional systems, the localization
 length $\xi_{\rm loc}$ is approximately equal to the
 mean free path \cite{Berezinskii73}. The equality between
 mean free path and localization length is valid even
 in the presence of an electric field, as the additional
  phase shift that the electron acquires as it moves along
  the field is canceled out on the return trip
  counter to the field \cite{Prigodin80}. Thus,
  the energy dependence of the backscattering
  probability of an individual scatterer gives
  rise to a position dependent localization
  length $\xi_{\rm loc}(x) \sim \ell (x)$.

Without applied external field, a localized electronic state centered
around a position $x_0$ is characterized by an exponentially decaying
wave function envelope $\Psi(x) \sim \exp(- |x - x_0|/\xi_{\rm
  loc})$. In the presence of an external field, the localization
length acquires a position dependence.  The electron kinetic energy
changes on a length ${\cal E}_0 /e_0 E$.  If the localization length
is much smaller than this length scale, $\xi_{\rm loc} \ll {\cal E}_0
/e_0 E$, the localization length varies slowly in space, and the wave
function envelope is given by \cite{Kirkpatrick86}
%
\begin{equation}
\Psi(x) \sim \left\{ \begin{array}{ll} e^{-(x_0 - x)/\xi_{\rm loc}(x_0)}
 & , x < x_0 \\
   e^{- \int_{x_0}^x d \tilde{x} \frac{1 }{ \xi_{\rm loc}(\tilde{x})}}
   & ,x > x_0
\end{array} \right. \ \ .
\end{equation}
%
Using the explicit position dependence of the mean free path given
in (\ref{meanfree.eq}), one finds that the envelope decays
asymptotically as a power
%
\begin{equation}
\Psi(x) \sim \left[  1 + \frac{(x - x_0)e E}{{\cal E}_0}
\right]^{-  \frac{m n_{\rm imp} u_0^2 }{ 2 e_0 E \hbar^2}}
\end{equation}
%
The state is only localized if it is normalizable, i.e. the
integral over the envelope squared
%
\begin{equation}
\int_{-\infty}^\infty d x \; \Psi^2 (x) < \infty
\end{equation}
%
must be finite. The wave function is only normalizable if the
envelope decays faster than $1/\sqrt{x}$ or if the electric field
is weaker than the critical electric field
%
\begin{equation}
E_c =  \frac{ m n_{\rm imp} u_0^2}{e_0 \hbar^2} \ .
\label{threshold.eq}
\end{equation}
%
This calculation  underestimates the exact result
\cite{Prigodin80} by a factor of two. We want to compare this
critical field to the crossover field
$E_0 = \frac{\pi v \hbar}{\xi_{\rm loc}^2 e_0}$
obtained from   Eq.~(\ref{eq:decay_rate}) for the
creep current. Using the relation
$\xi_{\rm loc} \approx \frac{ 2\hbar^2 {\cal E}_F}{m u_0^2 n_{\rm imp}}$
valid for noninteracting electrons,
the crossover
field Eq.~(\ref{threshold.eq}) can be rewritten as $E_c = E_0 k_F
\xi_{\rm loc}\gg E_0$. From this relation one sees that the
instanton mechanism is effective already at fields which are
parametrically smaller than the critical field for delocalization.

If the point scatterers are not of delta-function type but
characterized by a finite interaction range instead, even an
arbitrarily weak external electric field will lead to
delocalization on long length scales as the backscattering
probability drops significantly for electron wave vectors larger
than the inverse potential range.

How can this argument  be generalized to short range interacting
electrons? If the particle density is spatially homogeneous, the
kink kinetic energy far from the center of localization is much
larger than the Fermi energy of the interacting system and the kink
should essentially behave like a free particle. In this case, the
kink wave function would stay localized for fields below the
threshold value (\ref{threshold.eq}), and the power law tail of the
kink wave function could possibly enhance the tunneling current
derived in Sec. IV. Alternatively, the charge density may be locally
in equilibrium with the electro-chemical potential. In this
situation, both the particle  density and the Fermi wave vector are
position dependent. Assuming that the free electron relation ${\cal
E}_F = \frac{\hbar^2 k_F^2}{2 m}$ is at least approximately valid,
the Fermi wave vector acquires a position dependence
%
\begin{equation}
k_F (x)  = \sqrt{\frac{2 m}{\hbar^2} ({\cal E}_0 + e E x)} \ \ .
\end{equation}
%
Expressing the cutoff as $\Lambda = k_F$, the localization length
depends on the Fermi wave vector according to
%
\begin{equation}
\xi_{\rm loc} \propto  k_F^{-1} \left( k_F L_{\rm
FL}\right)^\frac{1}{(1 - K/K_c)} \ ,
\end{equation}
%
where the classical Fukuyama-Lee length $L_{\rm FL}$ is
independent of $k_F$. For a noninteracting system, $K/K_c= 2/3$
and one retrieves the dependence $\xi_{\rm loc} \sim k_F^2$.
For general $K$, the position dependence of the localization length is
\begin{equation}
\xi_{\rm loc}(x) \propto  \left[ \frac{2 m}{\hbar^2} ( {\cal E}_0
+ e E x)\right]^\frac{K }{2 (K_c -K)} \ L_{\rm FL}^\frac{ K_c}{K_c
- K} \ .
\end{equation}
%
Accordingly, the wave function envelope decays as a stretched
exponential $ \exp\left(- {\rm const}\; x^\frac{K_c - 3K/2}{ K_c
-K}\right)$ for $K < 1$ and does not decay to zero at all for $K
>1$. The argument leading to this result contains approximations
which may be correct only qualitatively. For this reason, we only
draw the conclusion that the delocalizing effect of an external
field is weakened by repulsive short range interactions and enhanced
by attractive ones. In order to determine whether the critical field
strength $E_c$ indeed has a discontinuity when varying $K$ across
one a more sophisticated calculation is necessary.

In conclusion, the delocalization mechanism studied in this
paragraph does  not seem to interfere with the creep current
described by formula Eq.~(4.25) for the case of repulsive short
range interactions.

\setcounter{equation}{0}
\section{Conclusion}\label{sec:conclusion}
In the present paper we extended Mott's and Shklovskii's approach for 1D
variable range hopping conductivity to short range interacting
electrons in charge density waves and Luttinger liquids using an
instanton approach. Following the recent paper \cite{ngd03}, we
calculated the quantum creep of charges at zero temperature and the
linear conductivity at finite temperatures for these systems. The
main results of the paper are equations (\ref{eq:decay_rate}) for
zero temperature and (\ref{eq:linear_conductivity_2}) for nonzero
temperatures, respectively.

In our approach  the quantum effects are weak but essential for the
transport. An applied electric field renders all classical ground
states metastable, and the motion between ground states along the
electric field occurs via quantum tunneling. We were interested in
the effect of the collective pinning by weak impurities, which
always localizes a 1D system with repulsive short range interactions
on a sufficiently large length scale. The presence of disorder
favors the charge transport: the corresponding pure system has an
excitation energy gap.

In the case with disorder, the notion of the materialization is
different from that in the pure case. The disordered system is
intrinsically inhomogeneous, so that the transport occurs as a
sequence of local materialization events. In the fermion picture
these correspond to separate hops whose positions and lengths
depend on the disorder configuration.  At zero temperature, the
typical hop length depends on the electric field $E$ and is given
by $L_x=\xi_{\rm loc}\sqrt{E_0/E}$. In this regime, the
conductivity reads: $\sigma\propto \exp(-\sqrt{E_0/E})$. Quantum
creep takes place for $E\ll E_0$. The results can be expressed in
terms of the localization length and the compressibility.

We compared the electric field $E_0$ with the threshold field
$E_c$ found in Refs.
\onlinecite{Prigodin80,Soukoulis+83,DeSiSo84,Kirkpatrick86,PriAl89}:
$E_0\sim E_c/(k_F \xi_{\rm loc})\ll E_c$, therefore we expect that
our creep conductivity result is valid.

We demonstrated that the inclusion of the dissipative term in the
action is necessary to find an unstable mode in the functional
integral. Besides, the dissipative bath takes into account the
possibility of inelastic processes and ensures the energy
conservation. The question about the energy balance does not
appear in the treatment of the pure system where the gain in the
electric field energy is spent on the increase of the kinetic
energy of departing kinks.

However, for weak dissipation, our result for the exponent in the
quantum creep regime  only weakly depends on the dimensionless
dissipative coefficient $\eta$. This feature can certainly change
as one crosses over to the nonzero temperature regime with linear
conductivity.

The effect of the Coulomb interaction in 1D was shown to be
irrelevant. It destroys the delocalization transition, but it is
not expected to change much in the phase already localized by the
disorder.

\acknowledgments The authors thank T. Giamarchi and D.G. Polyakov
for valuable comments on the manuscript and P. Le Doussal, S. Dusuel
and  S. Scheidl for useful discussions. S.M. acknowledges financial
support of DFG by SFB 608 and RFBR under grant No. 03-02-16173.

\appendix

\numberwithin{equation}{section}
\section{}
\label{append}
 Here we recall several relations for the bosonic representation of
 $1d$ spinless fermions. In a one-dimensional system of short range interacting
 fermions, the excitations can be understood as density fluctuations
 of bosonic nature. Following Ref. \onlinecite{KaneFisher}, we define
 the compressibility as the derivative of the particle density with
 respect to the chemical potential:
 $\kappa=\partial\rho/\partial\mu$. Correspondingly, the elastic
 energy density is given by $(\delta \rho)^{2}/(2\kappa)$.

Using expression (\ref{eq:n-zfr1}) for the long-wavelength
deviation of the density $\delta \rho=\partial_x\varphi/\pi$ from
$\rho_0$, we can rewrite the elastic energy in the form of the
second term in the Hamiltonian (\ref{eq:n-zfr6}) if
\begin{equation}\label{eq:app2}
    \kappa^{-1}=\frac{\pi\hbar v}{K}\,,\quad {\rm i.e.} \quad
    \frac{v}{K}=\frac{1}{\pi\hbar\kappa}\,.
\end{equation}
The last combination is usually denoted by $v_{N}$.

As both pressure and chemical potential depend only on the ratio
$N/V$, one can derive a relation between the conventional isothermal
compressibility \cite{AGD}
\begin{equation}\label{eq:app3}
   \tilde\kappa= -\frac{1}{V}\frac{\partial V}{\partial P}
= \rho^{-2}\frac{\partial \rho}{\partial\mu}=\frac{\kappa}{\rho^2}
\end{equation}
and the above defined $\kappa$.
We should note that he compressibility in Ref. \onlinecite{Haldane} was
defined as $\rho^{2}(\partial\mu/\partial\rho)$ which is the inverse
of the conventional thermodynamic isothermal compressibility
$\tilde{\kappa}$.

The kinetic energy is relevant for the description of dynamical
(quantum) phenomena. In a less rigorous way
we may argue that the first term
in Hamiltonian (\ref{eq:n-zfr6}) has its origin in a sum of
single-particle kinetic energies $p^2/(2m)$, where $p$ is the
momentum conjugate to the displacement, which is proportional to
$\varphi$, and $m$ is the effective mass. Then, using commutator
(\ref{eq:n-zfr3}) and the mentioned proportionality relations, we
obtain a relation between the coefficient in the Hamiltonian and
fermionic  quantities
\begin{equation}
    Kv=\frac{\pi\hbar\rho_{0}}{m}\,.
\end{equation}
This quantity is often referred to as $v_J$.

Thus, the parameters of the bosonized Hamiltonian, the stiffness
$K$ and the phason velocity, can be expressed in terms of the
fermion system parameters \cite{Haldane}:
\begin{equation}\label{eq:app6}
    K=\pi\hbar\sqrt{\frac{\rho_{0}\kappa}{m}}\,, \qquad
    v=\sqrt{\frac{\rho_{0}}{m\kappa}}\,.
\end{equation}
In the $g$-ology approach, the compressibility and the effective
mass are found as
\begin{equation}
\kappa=\frac{2\pi}{\pi\hbar(2\pi v_F +g_4+g_2)}, \qquad
m=\frac{2\pi p_F}{2\pi v_F + g_4 - g_2}\,,
\end{equation}
with the  Fermi momentum $p_F=\pi\hbar \rho_0$. Repulsion ($K<1$)
usually corresponds to a smaller compressibility and a larger
sound (phason) velocity.

\numberwithin{equation}{section}
\section{}
\label{appendb}

In this appendix, we justify our choice of an Ohmic dissipation in
Eq. (\ref{eq:n17}) and argue that it captures the generic features of
more general, and possibly material specific, dissipative mechanisms.
Coupling of the 1d electron systems to one- or higher dimensional
phonon degrees of freedom is the obvious candidate for the
realization of dissipation. This mechanism is extensively discussed in
the literature on Mott variable range hopping, see \cite{BoBr85} for
a review. The action for a bath
with arbitrary dispersion relation and coupling to the electron system
is given by
%
\begin{eqnarray}
{ S_{\rm dis} \over \hbar} & = & {1 \over  \Omega_{\rm ph}}
\sum_\alpha \int_{\beta/2}^{\beta/2}
d \tau \left[ {1 \over 2} c_\alpha(\tau)
\left(  -{\partial^2 \over \partial_\tau^2}  + \omega_\alpha^2 \right)
c_\alpha(\tau)  +  \right. \nonumber \\
& &
 \left.
 \int {d q\over 2 \pi} \varphi(-q,\tau) U(q) \Phi_\alpha(q) c_\alpha(\tau) \right] \ \ .
\end{eqnarray}
%
Here, the $c_\alpha$ are bosonic bath degrees of freedom with
eigenfunctions $\Phi_\alpha(q)$, eigenfrequencies $\omega_\alpha$,
 a coupling $U(q)$  to the electronic degrees of freedom, and $1/\Omega_{\rm ph}$ is 
 the volume element for one eigenstate in the $\alpha$--representation.
After integrating out the bath degrees of freedom, one finds the
dissipative term
%
\begin{align}
&{S_{\rm dis} \over \hbar}  =  -  T \sum_{\omega_n} \int \! {dq dq^\prime
\over (2 \pi)^2}
\varphi(q) K(q,q^\prime; \omega_n) \varphi(q^\prime)
\end{align}
%
with the kernel
%
\begin{eqnarray}
K(q,q^\prime; \omega_n)  =  { U(q) U(q^\prime)
  \over 2 \Omega_{\rm ph}} \sum_\alpha
{\Phi_\alpha(-q) \Phi_\alpha(- q^\prime) \over \omega_n^2 +
\omega_\alpha^2}\ .
\end{eqnarray}
%
The precise structure of the kernel $K$ depends on the dispersion
relation $\omega_\alpha$, the wave functions $\Phi_\alpha(q)$, and
the coupling $U(q)$  of the dissipative bath.
 For the instanton calculation in section IV we have
used a generic Ohmic dissipation $K(q,q^\prime; \omega_n) =
\eta \delta(q + q^\prime)| \omega_n|$ giving rise to a term
$\eta L_x \ln L_y$ in the instanton action. For a material specific choice of
parameters one may not find this Ohmic dissipation, but instead a
general dependence $\eta F(L_x,L_y)$. In the following, we will argue
that (i) for a weak coupling to the bath, one finds
only a logarithmic dependence  of the instanton action on the external electric field and hence no change of
the exponential field dependence in Eq.\ref{eq:decay_rate} and that (ii)
$S_{\rm dis}$ generally has a negative second derivative with respect to $L_y$.

For a weak coupling to the bath, the dominant process will be the hop
of a kink
accompanied by the emission of one excitation of the bath.
This process is
described by a first order Taylor expansion
%
\begin{eqnarray}
e^{- {\eta \over \hbar} F(L_x,L_y)}
\! \approx\! 1\! -\! {\eta \over \hbar} F(L_x,L_y) = 1\! -\! {\eta \over \hbar} e^{\ln F(L_x,L_y)} .
\end{eqnarray}
%
The zero order term does not make any contribution to the imaginary part of
the partition function $Z_1$.
In the spirit of such a Taylor expansion, one typically finds
that the preexponential factor for Mott variable range hopping conductivity
is proportional to the square of the
electron bath coupling and in our notation proportional to $\eta$.
According to this expansion, the instanton action contains a term $\ln F(L_x,L_y)$ and
depends via $L_x$, $L_y$ only in a
logarithmic fashion on the external electric field.

To illustrate our claim (ii) that generally ${\partial^2 F(L_x,L_y) \over \partial
L_y^2 } < 0$, we consider the specific example of a local coupling of the 1d
electron system to 1d acoustic phonons, i.e. the general index $\alpha$ corresponds
to momentum $q$, $U(q)\propto U_0 q^2$ (here $U_0$ is the dimensionless
electron-phonon coupling constant),
$\omega_\alpha = v_{\rm ph} q$. For an instanton with $L_x \ll {v_{\rm ph}
\over v} L_y$ one finds
%
\begin{eqnarray}
{S_{\rm dis} \over \hbar} \propto {\rm const. } - {L_x^2 \over L_y^2}
{U_0^2 v^2 \over v_{\rm ph}^2}
\end{eqnarray}
%
and hence a negative second derivative with respect to time.

\end{document}